\documentclass[sigconf, nonacm]{acmart}

\usepackage{graphicx}
\usepackage{tikz}
\usepackage{amsmath}
\usepackage{bm}
\usepackage{booktabs}
\usepackage{xcolor, pifont}
\usepackage{colortbl}
\usepackage{url}
\usepackage{multirow}
\usepackage{algorithm}
\usepackage{algpseudocode}
\usepackage{booktabs}
\usepackage{array} 
\usepackage{threeparttable}

\AtBeginDocument{%
  }

\definecolor{ctcol}{RGB}{200,30,30}
\definecolor{ptcol}{RGB}{30,100,200}
\definecolor{mixcol}{RGB}{120,60,160}
\colorlet{rowhl}{ptcol!12}  % light tint to mark the selected configuration

\definecolor{myblue}{RGB}{0, 102, 153}

\newcommand{\R}{\mathbb{R}}

\newcommand{\X}{\mathbf{X}}
\newcommand{\E}{\mathbf{E}}
\newcommand{\B}{\mathbf{B}}

\newcommand{\A}{\mathbf{A}}
\newcommand{\D}{\mathbf{D}}

\newcommand{\W}{\mathbf{W}}

\newcommand{\cmark}{\textcolor{green!70!black}{\ding{51}}}
\newcommand{\xmark}{\textcolor{red}{\ding{55}}}
\newcommand{\Enc}{\mathsf{Enc}}

\newtheorem{definition}{Definition}

\definecolor{myblue}{RGB}{0, 102, 153}

\begin{document}

\title{Privacy-Preserving GNN Inference on Financial Transaction Graphs via Homomorphic Encryption}


%% Anonymous submission — fill in real authors when de-anonymizing
\author{H. \c{C}a\u{g}r{\i} Bilgi}
\authornote{Both authors contributed equally to this research.}
\affiliation{%
  \institution{Delft University of Technology}
  \city{Delft}
  \country{Netherlands}}
\author{Kemal Mutluergil}
\authornotemark[1]
\affiliation{%
  \institution{Sabanc{\i} University}
  \city{Istanbul}
  \country{T\"urkiye}}
\author{Ceren Y{\i}ld{\i}r{\i}m}
\affiliation{%
  \institution{Sabanc{\i} University}
  \city{Istanbul}
  \country{T\"urkiye}}

\author{Kubilay Atasu}
\affiliation{%
  \institution{Delft University of Technology}
  \city{Delft}
  \country{Netherlands}}

\author{Erkay Savas}
\affiliation{%
  \institution{Sabanc{\i} University}
  \city{Istanbul}
  \country{T\"urkiye}}


\begin{abstract}
  Anti-money laundering (AML) in financial networks requires modeling transaction patterns that span multiple institutions. In practice, however, banks cannot share account and transaction features due to strict privacy regulations.
%  , leaving much of the graph signal inaccessible to any single party. 
  We propose a privacy-preserving graph neural network (GNN) inference framework tailored to this multi-bank setting. A central server holds the graph connectivity and pretrained model weights in plaintext, while each bank keeps its transaction features encrypted. Inference is then carried out entirely on encrypted features under the CKKS homomorphic encryption scheme, so that no plaintext transaction features or predictions are exposed to the server. The key to making this practical is an edge-centric GNN architecture, wherein message passing is expressed through operations on node--edge incidence matrices. This formulation not only preserves the edge-attributed multigraph structure of transaction networks but also reduces every aggregation step to plaintext--ciphertext matrix products. Since CKKS can evaluate only polynomial circuits, we also approximate each nonlinearity with a composite polynomial of the sign function, which we calibrate per activation site. On the IBM AML benchmark, our model achieves performance competitive with plaintext baselines, and with a consumer-grade GPU, classifies each queried transaction end-to-end in as little as \textbf{5.31} seconds, while per-transaction latency can be amortized to \textbf{1.65} seconds when optimized for throughput. 
  %while satisfying the constraints imposed by homomorphic encryption.
\end{abstract}

\begin{CCSXML}
<ccs2012>
   <concept>
       <concept_id>10010147.10010257.10010293.10010294</concept_id>
       <concept_desc>Computing methodologies~Neural networks</concept_desc>
       <concept_significance>500</concept_significance>
       </concept>
   <concept>
       <concept_id>10002978.10002979</concept_id>
       <concept_desc>Security and privacy~Cryptography</concept_desc>
       <concept_significance>500</concept_significance>
       </concept>
   <concept>
       <concept_id>10002978.10002991.10002995</concept_id>
       <concept_desc>Security and privacy~Privacy-preserving protocols</concept_desc>
       <concept_significance>300</concept_significance>
       </concept>
 </ccs2012>
\end{CCSXML}

\ccsdesc[500]{Computing methodologies~Neural networks}
\ccsdesc[500]{Security and privacy~Cryptography}
\ccsdesc[300]{Security and privacy~Privacy-preserving protocols}

\keywords{anti-money laundering, graph neural networks, homomorphic encryption, encrypted inference, financial transactions}



\maketitle

\section{Introduction}
%% ================================================================

Money laundering is the process by which illegally obtained funds are moved through financial transactions to disguise their origin. The global scale of the problem is substantial, with the UN estimating that up to \$2 trillion is laundered each year~\citep{unodc2022}. Laundering schemes rarely involve a single transfer; they typically unfold across chains of transactions, often spanning multiple banks. Detecting such schemes requires coordinated monitoring of the transactions.

Financial transaction data can be naturally represented as a graph, where each account is a node and each transaction between two accounts is an edge. Since multiple transactions can occur between the same pair of accounts, financial transaction graphs are multigraphs. Graph Neural Networks (GNNs) have emerged as a powerful tool for analyzing such graphs, and have been successfully applied to detect money laundering and related financial crimes in several recent works~\citep{weber2019anti, egressy2024provably, altman2023realistic, LaundroGraph, tam2019identifying, gnn_fraud_rel}. Their key strength is the ability to propagate information across the graph and capture suspicious patterns that span multiple hops between accounts.
However, in practice, no single financial institution has a complete view of the transaction graph. Money moves across many institutions, and each institution only has access to its own account and transaction features. Privacy regulations, such as GDPR, severely restrict the sharing of customer transaction records across institutions.

\begin{figure}[t!]
  \centering
  \includegraphics[width=1\linewidth]{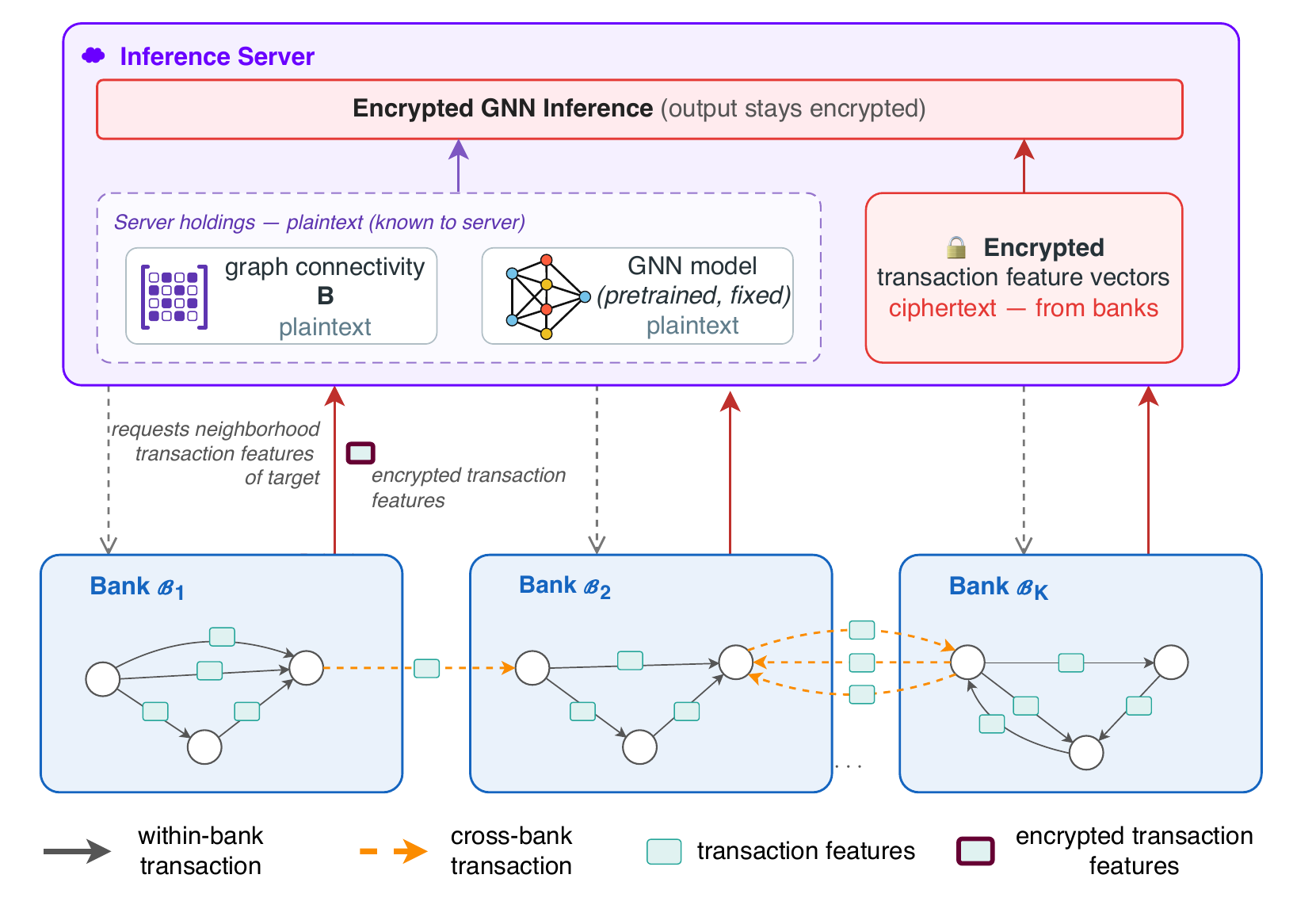}
  \caption{\textbf{Encrypted cross-bank GNN inference for AML.} The server holds graph connectivity $\mathbf{B}$ and pretrained weights $\mathbf{W}$ in plaintext; banks supply transaction features as ciphertexts. The GNN is evaluated under homomorphic encryption, leaving features and the prediction $\hat{\mathbf{Y}}$ encrypted to the server.}
  \label{fig:overview}
\end{figure}

One natural response to this fragmentation is federated learning~\citep{he2021fedgraphnn, yao2023fedgcn}, where banks collaboratively train a shared model without exchanging raw data. While this addresses collaborative model training, it does not by itself provide confidentiality of account and transaction features. Homomorphic encryption (HE) can address this challenge by computing directly on ciphertexts. In the setting we consider (Figure~\ref{fig:overview}), each bank encrypts its transaction features and sends them to a central inference server that holds the graph structure and model weights in plaintext. The server evaluates the GNN on the encrypted features and returns an encrypted prediction that only an authorized party can read. We assume the standard honest-but-curious adversary model and that no  feature or prediction is exposed to the server or other banks in plaintext.

Previous work has demonstrated the feasibility of HE-compatible GNN inference~\citep{ran2022cryptogcn, ran2023penguin, li2023lingcn, zhang2025ficgcn}, but none supports edge-attributed multigraphs, where attributes are attached to individual transactions and parallel edges between the same pair of accounts retain distinct identities. Moreover, all existing methods operate on a graph fixed and known at encryption time. In contrast, we perform per-query neighborhood sampling, extracting a fresh subgraph around queried transactions, which is crucial in an inductive setting~\citep{graphsage}. 
%exposing the model to diverse stochastically sampled neighborhoods rather than a fixed graph enables generalization to transactions and accounts unseen during training. 
To the best of our knowledge, ours is the first framework to support GNN inference on multigraphs with encrypted edge attributes as well as per-query stochastic neighborhood sampling on such graphs.

Our contribution is architectural: standard message-passing-based GNNs rely on scatter-gather 
operations, which are not naturally supported by HE arithmetic. On the other hand, adjacency-matrix-based formulations of such message-passing GNNs do not generalize to multigraphs with edge features, which can have several edges between the same two nodes. We address this challenge by introducing \emph{IncidenceGNN}, which exchanges messages using incidence matrix multiplications, preserving individual multigraph edges and their features while reducing the forward pass to matrix operations that can be efficiently mapped to HE arithmetic. The IncidenceGNN formulation therefore offers a more favorable accuracy–efficiency trade-off for encrypted edge classification. 

We overcome another challenge associated with the nonlinear nature of GNN activations: CKKS evaluates only polynomial circuits, and na\"ively approximating ReLU degrades the utility of the model. Prior HE-GNN work addresses this with simple quadratic approximations (Section ~\ref{sec:related}), which we find inadequate for the more complex decision boundaries of AML detection. Following~\citep{cheonetal2020, Lee2021PreciseAO, lee2022ppmlnn}, we replace each activation with a composite minimax polynomial approximation of the sign function, calibrated per activation site, and scheduling bootstrapping so that required approximation depth fits within the scheme's multiplicative budget. Testing on the IBM AML benchmark~\citep{altman2023realistic} via a consumer-grade GPU and no specialized hardware reveals that the resulting model is competitive with strong plaintext GNN baselines, and classifies each transaction query in a few seconds. To our knowledge, this is the first demonstration that transaction-level AML classification is practically achievable under full encryption at this scale, with near-zero accuracy loss.
%% ================================================================
\section{Related Work}
\label{sec:related}
%% ================================================================

% \paragraph{Homomorphic Encryption for GNNs.}
When combining fully homomorphic encryption with graph neural network inference, the dominant costs are ciphertext rotations and multiplicative depth. Among prior works in this area, CryptoGCN~\citep{ran2022cryptogcn} targets encrypted ST-GCN~\citep{st_gcn} inference under CKKS, utilizing a novel packing scheme and adjacency matrix decomposition method. Furthermore, it replaces ReLU with low-degree polynomial activations whose coefficients are trained jointly with the model. These activation sites are pruned at the layer level where the effect on accuracy is negligible via Activation Pruning~(AP). LinGCN~\citep{li2023lingcn} focuses on reducing the multiplicative depth, and proposes a ReLU approximation method that adapts node-wise trainable polynomials and distillation. FicGCN~\citep{zhang2025ficgcn} co-designs packing layout, per-layer aggregation mode, and node ordering against a CKKS cost model, reducing the rotation overhead of aggregation on encrypted features. Penguin~\citep{ran2023penguin} functions under a different threat model, treating the graph connectivity as private. In computation, Penguin interleaves several feature channels inside a single ciphertext so that the rotations required for aggregation are amortized across channels. DESIGN \citep{zhao2025design} assumes a similar threat model, and lets the server derive degree-based importance masks from an encrypted adjacency matrix to prune the graph and adapt activation-polynomial degree per node. Like the methods above, it assumes a single client uploads the entire graph in an adjacency-based encoding. Outside the FHE setting, OblivGNN~\citep{oblivgnn2024} and CryptGNN~\citep{cryptgnn2025} achieve secure GNN inference via multi-party computation and function secret
sharing rather than a single untrusted server; both target node-feature graphs
and, notably, CryptGNN identifies oblivious neighborhood sampling in the
MPC setting as expensive and instead performs it client-side, where the full graph
is already known~\citep{cryptgnn2025}, an escape unavailable when topology is
itself split across non-colluding banks. CryptoGCN, LinGCN, FicGCN, and Penguin (which supports inductive settings, but experiments are in a transductive setting) are tuned offline to a graph that is fixed at encryption time, and the resulting layout, decomposition, or schedule is amortized over repeated inferences on that same graph; evaluations are accordingly confined to a fixed 25-joint skeleton (CryptoGCN, LinGCN) and citation graphs of at most $\sim$20K nodes (Penguin, FicGCN). FicGCN's aggregation additionally bounds the number of neighbors summed per node, following the GraphSAGE formula~\citep{graphsage}, but this operates within that same fixed graph rather than varying the sampled subgraph from one query to the next. Our setting offers no such fixed graph to amortize over: a fresh neighborhood subgraph is sampled around each queried transaction, so the participating nodes and their connections change from one inference to the next.

These frameworks differ fundamentally in architecture and target task, so quantitative comparisons of computational and memory costs are not directly meaningful; Table~\ref{tab:qualitative_comparison} therefore provides a qualitative comparison with prior privacy-preserving GNN inference frameworks. To the best of our knowledge, our work is the first to enable encrypted edge classification on directed multi-graphs with query-based sampling. Query-based sampling restricts each encrypted inference to a relevant subgraph rather than requiring evaluation of the full graph, thereby reducing the memory overhead associated with large transaction networks. In addition, whereas all prior entries (except DESIGN which employs a multi-degree polynomial) approximate nonlinearities with quadratic polynomials, we employ a higher-degree composite approximation, \emph{AppReLU}, because low-degree polynomials degrade classification accuracy on datasets with complex decision boundaries (Section~\ref{subsec:apprelu}).

\begin{table*}[t]
\begin{threeparttable}
\centering
\caption{Qualitative comparison with prior work. Brackets ($[\cdot]$) denote encrypted data and the prime symbol ($'$) denotes the variable's state after query-based neighbor sampling. $\mathbf{A}/\mathbf{B}$: adjacency/incidence matrices, $\mathbf{X}/\mathbf{E}$: node/edge ft. matrices, $N/M$: number of graph nodes/edges, $d$: node/edge feature dimension, $b$: mini-batch size.}
\label{tab:qualitative_comparison}

\renewcommand{\arraystretch}{1.3}
\setlength{\tabcolsep}{7pt}
\footnotesize

\begin{tabular}{llp{3cm}lccl}
\toprule
\textbf{Work} &
\textbf{Threat model} &
\textbf{Downstream task} &
\textbf{\shortstack{Encrypted input\\domain}} &
\textbf{\shortstack{Neighbor\\sampling}} &
\textbf{\shortstack{Multi-graph /\\incidence}} &
\textbf{Nonlinearity} \\
\midrule

CryptoGCN~\cite{ran2022cryptogcn} &
$\mathbf{A}$,\ $[\mathbf{X}]$ &
Graph classification &
$[\mathbf{X}] \in \mathbb{R}^{b \times N \times d}$ &
\xmark &
\xmark &
Quadratic, layer-wise trainable \\ \relax

LinGCN~\cite{li2023lingcn} &
$\mathbf{A}$,\ $[\mathbf{X}]$ &
Graph / Node classification &
$[\mathbf{X}] \in \mathbb{R}^{N \times d}$ &
\xmark &
\xmark &
Quadratic, node-wise trainable\\ \relax

Penguin~\cite{ran2023penguin} &
$[\mathbf{A}]$, $[\mathbf{X}]$ &
Link prediction (GAE only) &
$[\mathbf{A}] \in \mathbb{R}^{N \times N}$,
$[\mathbf{X}] \in \mathbb{R}^{N \times d}$ &
\xmark &
\xmark &
Square ($x^2$) \\ \relax

FicGCN~\cite{zhang2025ficgcn} &
$\mathbf{A}$, $[\mathbf{X}]$ &
Graph classification / Link prediction (GAE only) &
$[\mathbf{X}] \in \mathbb{R}^{N \times d}$ &
\xmark \tnote{1} &
\xmark &
Square ($x^2$) for GAE\\

DESIGN~\cite{zhao2025design} &
$[\mathbf{A}]$, $[\mathbf{X}]$ &
Node classification &
$[\mathbf{A}] \in \mathbb{R}^{N \times N}$,
$[\mathbf{X}] \in \mathbb{R}^{N \times d}$ &
\xmark &
\xmark &
Adaptive, degree $\leq7$ poly. \\

\midrule

\cellcolor{rowhl}\textbf{Ours} &
\cellcolor{rowhl}$\mathbf{B}'$, $[\mathbf{E}']$ &
\cellcolor{rowhl}Edge classification &
\cellcolor{rowhl}$[\mathbf{E}'] \in \mathbb{R}^{M' \times d}$ &
\cellcolor{rowhl}\cmark &
\cellcolor{rowhl}\cmark &
\cellcolor{rowhl}AppReLU \\

\bottomrule
\end{tabular}
\begin{tablenotes}
\item[1] FicGCN's sampling bounds the number of neighbors per node, but every inference still processes a fixed graph known at encryption time; it does not perform per-query sampling.
\end{tablenotes}
\end{threeparttable}
\end{table*}

The IBM AML benchmark~\citep{altman2023realistic} provides synthetic transaction networks that closely follow real laundering patterns. Among the GNN-based methods proposed for this setting, Multi-GNN~\citep{egressy2024provably} extends standard GNNs with adaptations for directed multigraphs, which yields a provably more expressive architecture for detecting subgraph patterns. MEGA-GNN~\citep{megagnn} introduces a two-stage aggregation that first summarizes the edges between each pair of nodes before aggregating across neighbors. However, both architectures operate within the conventional node-centric message-passing paradigm and assume that the full transaction graph is available in one place. The closest work to our distributed setting is \citet{effendi2024aml}, who pursue the same collaborative, cross-bank AML goal under fully homomorphic encryption. Their GNN pipeline does not achieve encrypted inference, as the scatter--gather operator of message passing cannot be compiled to their FHE backend; the encrypted results they report come from an XGBoost model on a subset of the data. We overcome this obstacle with our \emph{IncidenceGNN}, whose message passing is formulated as incidence-matrix multiplications.

\section{Background}
\label{sec:background}
%% ================================================================

\subsection{Fully Homomorphic Encryption}

Homomorphic encryption allows computations to be carried out directly on encrypted data without the decryption key. Ciphertexts carry with them an inherent noise that grows with each homomorphic operation. Multiplications in particular consume levels of a finite \emph{multiplicative budget}; once exhausted, the ciphertext can no longer be decrypted correctly. Gentry \citep{gentry2009} removed this limit by \emph{bootstrapping}, which runs an encrypted version of the decryption algorithm over the noisy ciphertext to allow further computation. 

Popular FHE schemes include Cheon-Kim-Kim-Song scheme (CKKS)~\citep{CKKS2017}, Brakerski/Fan-Vercauteren scheme (BFV)~\citep{FV2012}, and Fast Fully Homomorphic Encryption over the Torus (TFHE)~\citep{TFHE}. The security of these schemes relies on the hardness assumption of Ring Learning with Errors (RLWE) \citep{regev2013}, which guarantees IND-CPA security. In addition, the BFV and CKKS schemes provide a feature known as \textit{batching}, allowing Single Instruction Multiple Data (SIMD) operations by encoding several messages into a single ciphertext.
In this work, we adopt the CKKS scheme due to its support for approximate arithmetic over floating-point numbers, which aligns well with our machine learning task. 

\subsection{Graph Notation}

A graph $\mathcal{G} = (\mathcal{V}, \mathcal{E})$ consists of a set of nodes $\mathcal{V}$ and a multiset of edges $\mathcal{E}$, with $N = |\mathcal{V}|$ and $M = |\mathcal{E}|$. In the directed case, each edge $e \in \mathcal{E}$ has a source node $u$ and target node $v$. The graph structure is commonly represented by the \emph{adjacency matrix} $\A \in \{0,1\}^{N \times N}$, where $A_{uv} = 1$ if an edge exists from $u$ to $v$ and $0$ otherwise.

In our financial transaction setting, the graph is a directed edge-attributed \emph{multigraph}: multiple distinct transactions may occur between the same pair of accounts, each with its feature vector. Since the adjacency matrix records only whether a connection exists, it cannot distinguish multiple edges between the same pair of nodes. The \emph{node--edge incidence matrix} dedicates a separate column to each edge, making it a natural representation for multigraphs.

\begin{definition}[Signed Node--Edge Incidence Matrix]
The \emph{signed node--edge incidence matrix} $\B \in \R^{N \times M}$ has rows indexed by nodes $v \in \mathcal{V}$ and columns indexed by edges $e \in \mathcal{E}$, with $B_{v,e} = -1$ if $v = \mathrm{src}(e)$, $B_{v,e} = +1$ if $v = \mathrm{dst}(e)$, and $B_{v,e} = 0$ otherwise.
% $e \in \mathcal{E}$, with entries:
% \begin{equation}
%   B_{v,e} =
%   \begin{cases}
%     -1 & \text{if } v \text{ is the source of } e, \\
%     +1 & \text{if } v \text{ is the target of } e, \\
%      0 & \text{otherwise.}
%   \end{cases}
% \end{equation}
\end{definition}
\begin{definition}[Source and Destination Incidence Matrices]
The \emph{source incidence matrix} $\B_{\mathrm{src}} \in \{0,1\}^{N \times M}$ and the \emph{destination incidence matrix} $\B_{\mathrm{dst}} \in \{0,1\}^{N \times M}$ are defined by $(\B_{\mathrm{src}})_{v,e} = 1$ iff $v = \mathrm{src}(e)$ and $(\B_{\mathrm{dst}})_{v,e} = 1$ iff $v = \mathrm{dst}(e)$.\end{definition}
% \begin{definition}[Source and Destination Incidence Matrices]
% The \emph{source incidence matrix} $\B_{\mathrm{src}} \in \{0,1\}^{N\times M}$ and the \emph{destination incidence matrix} $\B_{\mathrm{dst}} \in \{0,1\}^{N\times M}$ are defined by $(\B_{\mathrm{src}})_{v,e} = 1$ if and only if $v$ is the source of $e$, and $(\B_{\mathrm{dst}})_{v,e} = 1$ if and only if $v$ is the target of $e$. By construction, $\B = \B_{\mathrm{dst}} - \B_{\mathrm{src}}$.
% \end{definition}

Let each edge $e \in \mathcal{E}$ carry a feature vector $\mathbf{e} \in \R^d$, where $d \in \mathbb{N}$ is the feature dimension. These vectors are collected row-wise into the edge feature matrix $\E \in \R^{M \times d}$. Nodes (accounts) may also have associated features, collected into a node feature matrix $\X \in \R^{N \times d_v}$; however, the benchmark considered in this work contains no node features (Section~\ref{sec:experiments}). The task is \emph{edge classification}: predict a binary label $y_e \in \{0,1\}$ for each edge $e \in \mathcal{E}$.

\subsection{Graph Neural Networks}

Graph Neural Networks (GNNs) learn node (and edge) representations by iteratively aggregating information from local neighborhoods~\citep{mpnn}. At each layer $l$, the representation of node $v$ is updated:
\begin{equation}
    \mathbf{h}_v^{(l)} = \phi\!\left(\mathbf{h}_v^{(l-1)},\;
    \bigoplus_{u \in \mathcal{N}(v)} \psi\!\left(\mathbf{h}_u^{(l-1)}\right)\right),
\end{equation}
where $\mathcal{N}(v)$ is the neighborhood of $v$, $\psi$ is a learnable message function, $\bigoplus$ is a permutation-invariant aggregation (e.g., sum or mean), and $\phi$ is a learnable update function. With degree-normalized mean aggregation, this takes the matrix form $\mathbf{H}^{(l)} = \sigma\!\bigl(\hat{\A}\,\mathbf{H}^{(l-1)}\W^{(l)}\bigr)$~\citep{gcn}, where $\hat{\A} = \tilde{\D}^{-1/2}\tilde{\A}\tilde{\D}^{-1/2}$ is the symmetrically normalized adjacency matrix with self-loops.

\section{Problem Formulation}
\label{sec:problem}
%% ================================================================

% \paragraph{Multi-Bank Setting}

Consider a financial network observed across $K$ banks $\{\mathcal{B}_1, \ldots, \mathcal{B}_K\}$. Each bank $\mathcal{B}_k$ holds the records of the transactions that were made through its own accounts. In the graph, each transaction corresponds to an edge, and the set of edges under bank $k$ is denoted $\mathcal{E}_k \subseteq \mathcal{E}$, with $\bigcup_k \mathcal{E}_k = \mathcal{E}$. The corresponding transaction features are stored in $\E_k \in \R^{|\mathcal{E}_k| \times d}$ and are never revealed in plaintext to other banks or to the central server. A central \emph{inference server} holds the global graph structure: the full incidence matrices $\B, \B_{\mathrm{src}}, \B_{\mathrm{dst}} \in \R^{N \times M}$ and a trained model with plaintext weights. As is standard in privacy-preserving inference, the model is trained in plaintext and only inference is performed on encrypted inputs~\cite{ran2022cryptogcn,effendi2024aml}. Account identifiers are pseudonymized. The inference server seeks to classify transaction edges without accessing plaintext feature data.

% \paragraph{Inference Protocol}
\label{sec:inference_protocol}

Real-world transaction graphs contain millions of edges, making full-graph inference intractable. Following~\cite{graphsage}, we perform plaintext neighborhood sampling around each query edge, reducing inference to a small subgraph. Given the subgraph $(\mathcal{V}', \mathcal{E}')$, with incidence matrix $\B' \in \R^{N' \times M'}$, the inference protocol is:

\begin{enumerate}  
  \item \textbf{Edge Query.} An initiator bank raises suspicion about one of its transactions and sends a query to the central server for that edge to be evaluated with the server's model.
  \item \textbf{Graph sampling.} The server obtains a subgraph $(\mathcal{V}', \mathcal{E}')$, with incidence matrix $\B' \in \R^{N' \times M'}$ via neighborhood sampling on the queried edge, using plaintext graph. 
  \item \textbf{Encrypted feature request.} The server requests an encrypted feature vector $\Enc(\mathsf{pk}, \mathbf{e})$, for each edge $e \in \mathcal{E}'$ (and $\Enc(\mathsf{pk}, \mathbf{x}_v)$ for each node $v \in \mathcal{V}'$, where
account attributes are available) from its host bank $\mathcal{B}_k$, where $\mathsf{pk}$ is the public key of the initiator bank.
  \item \textbf{Encrypted inference.} The server evaluates the GNN model on the encrypted feature matrix $\Enc(\mathsf{pk}, \E')$ using plaintext incidence matrices and model weights. This produces encrypted logits.
  \item \textbf{Decryption.} Encrypted logits are returned to the initiator bank which decrypts them with the secret key $\mathsf{sk}$.
\end{enumerate}

%% ================================================================
\section{Method}
\label{sec:method}
%% ================================================================

\subsection{Message Passing with Incidence Matrices for Directed Edge-attributed Multigraphs} \label{sec:incidenceGNN}

\begin{figure}[h]
  \centering
  \includegraphics[width=0.9\linewidth]{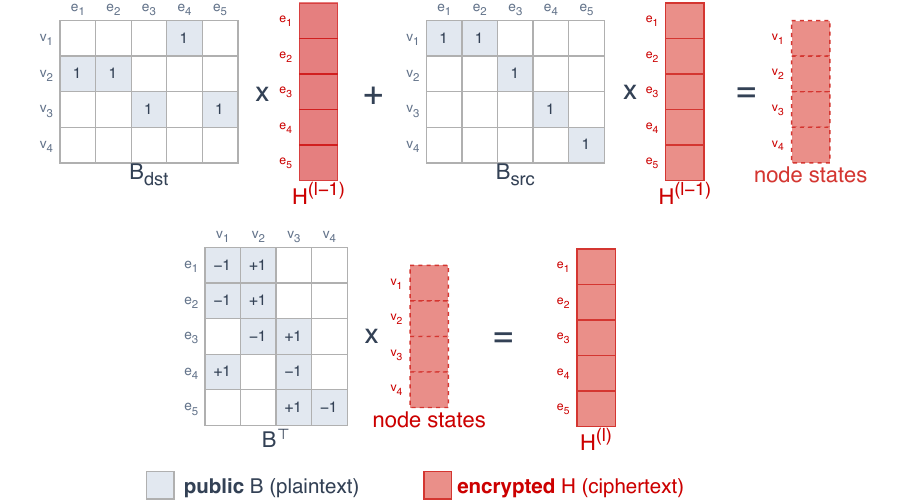}
  \caption{One layer of incidence matrix based edge-centric message-passing. Weight matrices are omitted for simplicity.}
  \vspace{-8pt}
  \label{fig:incidence_gnn}
\end{figure}

We adopt an edge-centric formulation of message passing for transaction graphs: transactions are the primary entities, each retaining its own feature vector, and each layer applies an edge-to-node transformation followed by a node-to-edge transformation, as shown in Figure~\ref{fig:incidence_gnn}. Our setting imposes two requirements on the message-passing implementation. First, the graph is a multigraph, so each edge must be tracked individually. Second, the computation must reduce to dense matrix multiplications, since HE cannot support the irregular scatter-gather operations that data-dependent control flow requires. Neither standard implementation satisfies both. Libraries such as PyTorch Geometric~\citep{Fey2019} handle multigraphs naturally via an edge-index, but rely on scatter-gather and are therefore HE-incompatible. The GCN-like adjacency-matrix form is HE-friendly, but $\hat{\A}$ can't represent individual edges in multigraphs.

We satisfy both constraints by reformulating message passing with the incidence matrices $\B_{\mathrm{dst}}$ and $\B_{\mathrm{src}}$: each edge occupies its own column, and aggregation reduces to the matrix products $\B_{\mathrm{dst}}\E$ and $\B_{\mathrm{src}}\E$. Keeping the two paths separate lets each account aggregate its incoming and outgoing transactions independently, a common choice in  directed GNNs~\citep{edge_dir,egressy2024provably}. The resulting account representations are then propagated back to edges through $\B^{\top}$. The procedure is summarized in Algorithm~\ref{alg:incidencegnn-full}. Furthermore, in practice we use degree-normalized incidence matrices $\hat{\B}_{(\cdot)} = \D^{-1/2}\,\B_{(\cdot)}$ in place of $\B_{\mathrm{dst}}$, $\B_{\mathrm{src}}$, and $\B$, where $\D$ is the diagonal matrix of the node degrees (in-, out-, and total degree, respectively); since the graph structure is held in plaintext, this normalization incurs no homomorphic cost. Account attributes, where available, can be incorporated directly into the node update: $\X$ is node-indexed and enters the node state as an additional linear term (line 10, Algorithm~\ref{alg:incidencegnn-full}).

\begin{algorithm}[h]
\caption{IncidenceGNN}
\label{alg:incidencegnn-full}
\begin{flushleft}
\textbf{Inputs:} Edge features $\E \in \R^{M \times d}$; optional node features
$\X \in \R^{N \times d_v}$; incidence matrices $\B_{\mathrm{dst}}, \B_{\mathrm{src}} \in \{0, 1\}^{N \times M}, \B \in \{-1,0,1\}^{N \times M}$;
weights $\mathbf{W}_{\mathrm{enc}}$, $\{\mathbf{W}^{(l)}_{\mathrm{fwd}}, \mathbf{W}^{(l)}_{\mathrm{rev}}, \mathbf{W}^{(l)}_{\mathrm{lin}}\}_{l=1}^{L}$, $\mathbf{W}_{\mathrm{cls0}}, \mathbf{W}_{\mathrm{cls1}}$ with biases $\mathbf{b}_{(\cdot)}$; batch norms $\mathrm{BN}$; activation $\sigma$.\\
\textbf{Output:} Edge logits $\mathbf{Y}$.
\end{flushleft}
\begin{algorithmic}[1]

\State $\mathbf{H}_{\mathrm{edge}}^{(0)} \gets \sigma\!\bigl(\mathrm{BN}_{\mathrm{enc}}(\E\,\mathbf{W}_{\mathrm{enc}}^{\top} + \mathbf{b}_{\mathrm{enc}})\bigr)$ \Comment{edge encoding}

\If{node features $\X$ are available}
    \State $\mathbf{H}^{(0)}_{\mathrm{node}}
    \gets
    \sigma\!\bigl(
    \mathrm{BN}_{\mathrm{enc}}^{v}
    (\X\mathbf{W}_{\mathrm{enc}}^{v\top}
    +\mathbf{b}_{\mathrm{enc}}^{v})
    \bigr)$
    \Comment{node encoding}
\Else
    \State $\mathbf{H}^{(0)}_{\mathrm{node}}\gets \mathbf{0}$
\EndIf

\For{$l = 1$ \textbf{to} $L$}
  \State $\mathbf{H}^{(l)}_{\mathrm{in}} \gets \sigma\!\bigl(\mathrm{BN}^{(l)}_{\mathrm{fwd}}((\hat{\B}_{\mathrm{dst}}\,\mathbf{H}_{\mathrm{edge}}^{(l-1)})\,\mathbf{W}_{\mathrm{fwd}}^{(l)\top} + \mathbf{b}^{(l)}_{\mathrm{fwd}})\bigr)$ \Comment{incoming}
  
  \State $\mathbf{H}^{(l)}_{\mathrm{out}} \gets \sigma\!\bigl(\mathrm{BN}^{(l)}_{\mathrm{rev}}((\hat{\B}_{\mathrm{src}}\,\mathbf{H}_{\mathrm{edge}}^{(l-1)})\,\mathbf{W}_{\mathrm{rev}}^{(l)\top} + \mathbf{b}^{(l)}_{\mathrm{rev}})\bigr)$ \Comment{outgoing}

    \State $\mathbf{H}^{(l)}_{\mathrm{node}}
    \gets
    \mathbf{H}^{(l-1)}_{\mathrm{node}}
    +
    \mathbf{H}^{(l)}_{\mathrm{in}}
    +
    \mathbf{H}^{(l)}_{\mathrm{out}}$
  
  \State $\mathbf{M}^{(l)} \gets \hat{\B}^{\top}\,\mathbf{H}^{(l)}_{\mathrm{node}}$ \Comment{node $\to$ edge}
  \State $\mathbf{H}_{\mathrm{edge}}^{(l)} \gets \sigma\!\bigl(\mathrm{BN}^{(l)}_{\mathrm{edge}}(\mathbf{M}^{(l)}\,\mathbf{W}_{\mathrm{lin}}^{(l)\top} + \mathbf{b}^{(l)}_{\mathrm{lin}})\bigr)$ \Comment{edge update}
\EndFor
\State $\mathbf{A} \gets \sigma\!\bigl(\mathbf{H}_{\mathrm{edge}}^{(L)}\,\mathbf{W}_{\mathrm{cls0}}^{\top} + \mathbf{b}_{\mathrm{cls0}}\bigr)$
\State $\mathbf{Y} \gets \mathbf{A}\,\mathbf{W}_{\mathrm{cls1}}^{\top} + \mathbf{b}_{\mathrm{cls1}}$ \Comment{classification head}
\State \Return $\mathbf{Y}$
\end{algorithmic}
\end{algorithm}
% \vspace{-4pt}

\subsection{Homomorphic Evaluation Methodology}
\label{sec:fhe_methodology}
FHE mandates that computations be expressed entirely as additions, multiplications, and ciphertext rotations. As formulated in Algorithm~\ref{alg:incidencegnn-full}, our model consists solely of a series of matrix operations. The only non-arithmetic components are the ReLU activations, which we replace with polynomial approximations (Section~\ref{subsec:apprelu}), and the batch-normalization layers, which are folded into the preceding linear maps \cite{bnfolding}.

\subsubsection{Efficient Homomorphic Matrix Multiplication}
\label{subsubsec:fhe_matmul}

Evaluating the dense matrix operations defined in Algorithm~\ref{alg:incidencegnn-full} requires an encoding scheme that is (i) computationally efficient, (ii) produces outputs that chain across layers without repacking, (iii) handles left-multiplications by plaintext matrices naturally, and (iv) scales to matrices whose dimensions exceed the slot capacity of a single ciphertext. We adopt the Tricycle framework~\cite{tricycle2025}, a generalization of Bicyclic encoding ~\cite{bicyclic2025} which (i) reduces rotation cost from $O(n)$ to $O(\sqrt{n})$ via a baby-step giant-step sweep, (ii) lets ciphertext outputs chain directly across layers without repacking, (iii) supports our left-multiplications $\mathbf{B}\mathbf{H}$ at zero extra cost by encoding $\mathbf{H}$ and $\mathbf{H}^\top$ identically, and (iv) scales to matrices exceeding slot capacity via block partitioning.

Since each bank encrypts its features independently, incoming ciphertexts arrive in a row-wise layout rather than the joint bicyclic layout our matrix multiplications (MatMul) require; the server converts between them via a one-time rotate-and-add packing step and a fixed transformation matrix \cite{bicyclic2025}. 

% This conversion is a per-query preprocessing cost not included in the latency measurements of Table~\ref{tab:fhe_latency}.

\subsubsection{Homomorphic ReLU Approximation via AppReLU}
\label{subsec:apprelu}

Since CKKS only evaluates polynomial circuits, we approximate $\mathrm{ReLU}(x)=\mathrm{max}(0, x)$ via the identity $\mathrm{ReLU}(x)=x(1+\mathrm{sgn}(x))/2$ replacing $\mathrm{sgn}$ with the composite minimax polynomial family of \citep{cheonetal2020}, following \citep{Lee2021PreciseAO, lee2022ppmlnn}. We use the degree-9 polynomial

\begin{equation}
    f_4(t) \;=\; \frac{t}{128}\bigl(315 - 420t^2 + 378t^4 - 180t^6 + 35t^8\bigr),
    \label{eq:f4}
\end{equation}
which maps $[-1,1] \to [-1,1]$ and closely approximates $\mathrm{sgn}(t)$ on that interval. Composing $f_4$ with itself $d$ times improves the sign approximation at the cost of additional multiplicative depth (4 levels per composition via Paterson-Stockmeyer)

Following \cite{Lee2021PreciseAO, lee2022ppmlnn}, given an encrypted activation $x$ and a clipping bound $B > 0$, one per activation site of the network, we define
\begin{equation}
    \mathrm{AppReLU}(x;\, B, d) \;=\; \frac{x}{2}\Bigl(1 + f_4^d\!\Bigl(\frac{x}{B}\Bigr)\Bigr).
    \label{eq:apprelu}
\end{equation}
where dividing by $B$ rescales inputs into $f_4$'s accurate range and the outer scaling implements the $\mathrm{ReLU}$ half-wave rectification, all with plaintext constants.

Each activation site requires multiple nontrivial polynomial evaluations, quickly exhausting the ciphertext's multiplicative budget; we use \texttt{slim\_bootstrapping} ~\cite{slimbootstrapping}
to refresh it. Under our CKKS parameterization, one bootstrap call covers only a single $f_4$ evaluation, so the circuit performs 8 external bootstraps (one per activation site) plus $8d$ internal bootstraps (one per composition), for $8 + 8d = 48$ total calls at $d = 5$ --- the dominant cost in Table ~\ref{tab:fhe_latency}.

% \begin{algorithm}[h]
% \caption{$\mathsf{evalAppReLU}$: homomorphic AppReLU evaluation}
% \label{alg:apprelu}
% \begin{algorithmic}[1]
% \Require Ciphertext $\mathsf{ct}$ at the output level of \texttt{slim\_bootstrapping};
%          bound $B > 0$; depth $d$; threshold $L_\mathrm{thr} = L_\mathrm{min} + 4$;
%          bootstrap key $\mathsf{bk}$; relinearization key $\mathsf{rlk}$
% \State $\mathsf{ct}_x \gets \mathsf{ct}$   
% \State $\mathsf{ct}_t \gets \mathsf{ct} \otimes (1/B)$
% \For{$i \gets 1$ \textbf{to} $d$}
%     \State $\mathsf{ct}_t \gets \mathsf{evalPoly}(f_4,\, \mathsf{ct}_t,\, \mathsf{rlk})$
%    \Comment{degree-9 Paterson--Stockmeyer; costs 4 levels}
%     \If{$\mathsf{ct}_t.\mathsf{level}() < L_\mathrm{thr}$}
%         \State $\mathsf{ct}_t \gets \mathsf{slim\_bootstrapping}(\mathsf{ct}_t,\, \mathsf{bk},\, \mathsf{rlk})$
%         \Comment{internal bootstrap; fires after every composition under $L=26$}
%     \EndIf
% \EndFor
% \State $\mathsf{ct}_\mathrm{sum} \gets \mathsf{ct}_t \oplus 1$            \Comment{$1 + f_4^d(x/B)$}
% \State $\mathsf{ct}_\mathrm{out} \gets \mathsf{ct}_x \otimes_\mathsf{ct} \mathsf{ct}_\mathrm{sum}$
% \Comment{$x \cdot (1 + f_4^d(x/B))$; }
% \State $\mathsf{ct}_\mathrm{out} \gets \mathsf{ct}_\mathrm{out} \otimes 0.5$
% \State \Return $\mathsf{ct}_\mathrm{out}$
% \end{algorithmic}
% \end{algorithm}

\paragraph{Knowledge-distillation finetuning.}
To adapt the ReLU-trained network to AppReLU's approximation error, we finetune with App\-ReLU substituted for ReLU. Since na\"ive cross-entropy finetuning degrades accuracy (Section ~\ref{sec:experiments}), we instead apply a knowledge-distillation objective
\begin{equation}
    \mathcal{L} = \alpha T^2 \cdot \mathrm{KL}\!\left(\frac{p_\text{teacher}}{T}\,\Big\|\,
    \frac{p_\text{student}}{T}\right) + (1-\alpha)\,\mathrm{CE},
    \label{eq:kd_loss}
\end{equation}
with temperature $T = 3$, weight $\alpha = 0.7$, and the original ReLU network as
the frozen teacher, initializing the student from the teacher's weights.
We quantify the effect of the finetuning objective in the ablation
(Table~\ref{tab:kd_ablation}, Section~\ref{sec:encrypted_inference}).

\section{Experiments}
\label{sec:experiments}
%% ================================================================

In this section, we evaluate IncidenceGNN in plaintext against existing graph-based AML detection models (Section~\ref{subsec:plain_text_exp}), then present ablations for architectural and hyperparameter choices. Finally, we measure the latency of privacy-preserving inference, and assess fidelity of FHE predictions (Section~\ref{sec:encrypted_inference}).

\paragraph{Datasets and setup.} We evaluate on the \textbf{IBM AML} benchmark~\citep{altman2023realistic}, publicly available synthetic anti-money laundering datasets designed to capture realistic transaction characteristics. We consider four dataset variants: \textbf{Small} and \textbf{Medium} graphs, each provided with a higher-illicit (\textbf{HI}) and lower-illicit (\textbf{LI}) split that differ in the prevalence of laundering transactions. Following prior work~\cite{egressy2024provably}, we use a temporal train/validation/test split. Each dataset is a directed edge-attributed multigraph, with nodes representing accounts and edges representing transactions. Each transaction contains four attributes: timestamp, amount, currency, and payment format. Accounts have no associated features ($\X=\mathbf{0}$). The task is binary edge classification, where the goal is to identify transactions belonging to money laundering patterns. The laundering class is the minority class, with LI variants substantially more imbalanced than HI variants. We report minority-class F1 and PR-AUC, averaged over at least five random seeds. We use minority-class F1 as the primary threshold-dependent metric, while PR-AUC provides a threshold-independent measure of ranking quality. Plaintext models are evaluated with an inference batch size of $8{,}192$ and use ReLU as the non-linear activation function.

% We further compare against three adjacency-based GINE variants~\citep{xu2018gin} in the ablation study of Section~\ref{sec:ablations}. These comparisons isolate the effect of graph representation: incidence- versus adjacency-based message passing.

\subsection{Plaintext Results}\label{subsec:plain_text_exp}

Table~\ref{tab:plaintext} compares IncidenceGNN against the AML detection models of~\citet{egressy2024provably}, in plaintext. IncidenceGNN outperforms the standard GIN baseline on three of the four splits and matches it on Medium-LI within variance, while implementing sum aggregation entirely through sparse incidence-matrix multiplication. Against the stronger Multi-GIN baseline, IncidenceGNN remains competitive across most splits, matching or slightly outperforming it on Small-HI within variance and achieving comparable performance on Small-LI. The main gap appears on Medium-LI, where IncidenceGNN underperforms Multi-GIN. However, Table~\ref{tab:he_cost} highlights that plaintext accuracy alone does not capture the relevant trade-off under homomorphic inference: IncidenceGNN requires fewer nonlinear activations, performs aggregation through sparse matrix multiplication, and avoids scatter–gather operations and feature concatenation, resulting in a $2.61\times$ speedup over Multi-GIN. The incidence-based formulation therefore offers a more favorable accuracy–efficiency trade-off for encrypted edge classification.

\begin{table}[h]
  \centering
  \caption{Plaintext minority-class F1 (\%) on four IBM AML splits.}
  \label{tab:plaintext}
  \resizebox{\columnwidth}{!}{%
  \begin{tabular}{lccccc}
    \toprule
    Model  & Small-HI & Small-LI & Medium-HI & Medium-LI \\
    \midrule
    % GIN~\cite{egressy2024provably} & $47.73 \pm 7.86$ & $20.62 \pm 2.41$ & $49.26 \pm 4.02$ & $6.19 \pm 8.32$ \\
    % Multi-GIN~\cite{egressy2024provably}  & $64.79 \pm 1.22$ & $26.88 \pm 6.63$ & $58.92 \pm 1.83$ & $16.30 \pm 4.73$ \\
    GIN        & $46.50 \pm 4.11$ & $19.93 \pm 3.55$ & $58.65 \pm 2.50$ & $25.36 \pm 1.49$ \\
    Multi-GIN  & $62.66 \pm 1.73$ & $32.21 \pm 0.99$ & $67.72 \pm 0.94$ & $31.24 \pm  2.12$ \\
    \midrule
    \rowcolor{rowhl}
    IncidenceGNN (ours) & $63.72 \pm 1.86$ & $30.12 \pm 2.08$ & $63.20 \pm 1.38$ & $24.52 \pm 3.42$ \\
    \bottomrule
  \end{tabular}
  }
  \vspace{1em}

\caption{Architectural and runtime cost per model (2 layers, hidden dim $64$). \textbf{MatMul-only}: no scatter--gather; \textbf{\# Concat}: concatenations forcing ciphertext repacking.}
\label{tab:he_cost}
\resizebox{\columnwidth}{!}{%
\begin{tabular}{lcccc}
  \toprule
  Model & \# Nonlinear activations & MatMul-only & \# Concat & Time/batch (ms) \\
  \midrule
  GIN        & $11$ & \xmark & $4$ & $84.48$   \\
  Multi-GIN  & $17$ & \xmark & $6$ & $231.43$  \\
  \midrule
  \rowcolor{rowhl}
  IncidenceGNN (ours)       & $8$  & \cmark & $0$ & $88.61$  \\
  \bottomrule
\end{tabular}
}

\end{table}
 
% We find that the Medium-LI degradation is closely tied to ego and port numbers. Removing these features improves minority-class F1 from 24.52 to 35.64 (Table~\ref{tab:adjacency_baseline}), while retaining them remains beneficial on the other datasets.

% \subsubsection{Ablation Studies of Plaintext Model} \label{sec:ablations}
% We ablate our model along two axis: the architectural and hyperparameter design choices (Table~\ref{tab:design_choices}), and the incidence-based formulation against the adjacency-based message passing (Table~\ref{tab:adjacency_baseline}).

\paragraph{Ablation of architectural choices and hyperparameters.} Table~\ref{tab:design_choices} varies one design factor at a time. The message-passing path has the largest impact: using only forward or reverse messages substantially degrades performance, showing that  incoming and outgoing messages provide complementary laundering signals. Shared weights perform poorly, indicating that forward and reverse messages require independent parameterization and supporting the two-incidence design ($\B_{\mathrm{dst}}, \B_{\mathrm{src}}$). Removing edge direction reduces performance, so we retain the directional form. The \emph{multigraph adaptations} from~\citet{egressy2024provably} (ego IDs and port numbering) provide consistent gains and incur negligible encrypted overhead, as they are derived from the plaintext graph structure and require only plaintext--ciphertext additions. We therefore include them in the baseline. Increasing the hidden dimension to $128$ offers only marginal gains while increasing ciphertext costs, so we use $h=64$.

\begin{table}[t]
  \centering
  \caption{Design and hyperparameter ablations on IBM AML (Small-HI, Small-LI). Each block varies one factor of the baseline; the selected setting is \colorbox{rowhl}{highlighted}. Minority-class test F1 and PR-AUC (\%), mean $\pm$ std over seeds.}
  \label{tab:design_choices}
  \setlength{\tabcolsep}{3pt}
  \renewcommand{\arraystretch}{1.15}
  \resizebox{\columnwidth}{!}{%
  \begin{tabular}{llcccc}
    \toprule
    & & \multicolumn{2}{c}{Small-HI} & \multicolumn{2}{c}{Small-LI} \\
    \cmidrule(lr){3-4}\cmidrule(lr){5-6}
    & Variant & F1 & PR-AUC & F1 & PR-AUC \\
    \midrule
    \multirow{4}{*}{\rotatebox[origin=c]{90}{Message path}}
      & \cellcolor{rowhl}Forward + reverse (selected) & \cellcolor{rowhl}$63.72{\scriptstyle\pm 1.86}$ & \cellcolor{rowhl}$62.15{\scriptstyle\pm 1.78}$ & \cellcolor{rowhl}$30.12{\scriptstyle\pm 2.08}$ & \cellcolor{rowhl}$24.74{\scriptstyle\pm 1.53}$ \\
      & Forward-only             & $43.07{\scriptstyle\pm 3.12}$ & $38.58{\scriptstyle\pm 3.52}$ & $11.18{\scriptstyle\pm 4.95}$ & $6.77{\scriptstyle\pm 3.45}$ \\
      & Reverse-only             & $57.96{\scriptstyle\pm 3.22}$ & $54.84{\scriptstyle\pm 3.32}$ & $21.23{\scriptstyle\pm 2.61}$ & $15.62{\scriptstyle\pm 2.86}$ \\
      & Shared weights           & $21.37{\scriptstyle\pm 8.36}$ & $15.35{\scriptstyle\pm 8.78}$ & $12.86{\scriptstyle\pm 3.22}$ & $8.49{\scriptstyle\pm 3.65}$ \\
    \midrule
    \multirow{2}{*}{\rotatebox[origin=c]{90}{Incid.}}
      & \cellcolor{rowhl}Directional (selected)       & \cellcolor{rowhl}$63.72{\scriptstyle\pm 1.86}$ & \cellcolor{rowhl}$62.15{\scriptstyle\pm 1.78}$ & \cellcolor{rowhl}$30.12{\scriptstyle\pm 2.08}$ & \cellcolor{rowhl}$24.74{\scriptstyle\pm 1.53}$ \\
      & Undirected               & $56.26{\scriptstyle\pm 3.01}$ & $53.01{\scriptstyle\pm 2.72}$ & $30.41{\scriptstyle\pm 0.96}$ & $16.62{\scriptstyle\pm 1.99}$ \\
    \midrule
    \multirow{4}{*}{\rotatebox[origin=c]{90}{Multi. adap.}}
      & None                     & $61.88{\scriptstyle\pm 1.92}$ & $59.18{\scriptstyle\pm 2.81}$ & $26.90{\scriptstyle\pm 0.72}$ & $21.65{\scriptstyle\pm 0.82}$ \\
      & + Ego IDs                & $63.47{\scriptstyle\pm 2.76}$ & $62.69{\scriptstyle\pm 1.92}$ & $29.98{\scriptstyle\pm 2.20}$ & $24.18{\scriptstyle\pm 2.20}$ \\
      & + Port IDs               & $62.21{\scriptstyle\pm 2.24}$ & $59.74{\scriptstyle\pm 3.57}$ & $27.66{\scriptstyle\pm 1.81}$ & $22.38{\scriptstyle\pm 1.05}$ \\
      & \cellcolor{rowhl}+ Ego + Ports (selected)  & \cellcolor{rowhl}$63.72{\scriptstyle\pm 1.86}$ & \cellcolor{rowhl}$62.15{\scriptstyle\pm 1.78}$ & \cellcolor{rowhl}$30.12{\scriptstyle\pm 2.08}$ & \cellcolor{rowhl}$24.74{\scriptstyle\pm 1.53}$ \\
    \midrule
    \multirow{4}{*}{\rotatebox[origin=c]{90}{Hidden dim}}
      & $16$                     & $57.20{\scriptstyle\pm 1.78}$ & $56.12{\scriptstyle\pm 3.42}$ & $25.32{\scriptstyle\pm 1.43}$ & $18.28{\scriptstyle\pm 1.06}$ \\
      & $32$                     & $60.35{\scriptstyle\pm 1.35}$ & $57.37{\scriptstyle\pm 1.92}$ & $28.42{\scriptstyle\pm 4.02}$ & $21.98{\scriptstyle\pm 4.04}$ \\
      & \cellcolor{rowhl}$64$ (selected)              & \cellcolor{rowhl}$63.72{\scriptstyle\pm 1.86}$ & \cellcolor{rowhl}$62.15{\scriptstyle\pm 1.78}$ & \cellcolor{rowhl}$30.12{\scriptstyle\pm 2.08}$ & \cellcolor{rowhl}$24.74{\scriptstyle\pm 1.53}$ \\
      & $128$                    & $65.09{\scriptstyle\pm 0.88}$ & $62.74{\scriptstyle\pm 1.10}$ & $28.86{\scriptstyle\pm 1.77}$ & $24.03{\scriptstyle\pm 0.62}$ \\
    \bottomrule
  \end{tabular}}
\end{table}

\subsection{Encrypted Inference}
\label{sec:encrypted_inference}

\paragraph{Setup.} 

We evaluate FHE inference on the Small-HI test set using $N = 65536$ as the polynomial degree, scale $\Delta = 2^{50}$, and $|Q| = 27$ primes in the modulus chain ($\log_2{\prod_{i=0}^{26}q_i} \approx 60 + 26 \times 50 = 1360$ bits) and $|P| = 5$ primes in the auxiliary chain for key switching operations, achieving 128-bit classical security. We implement the protocol using the HEonGPU \citep{heongpu} library and run on an NVIDIA RTX 5090 GPU, equipped in a system with dual Intel Xeon Silver 4514Y CPUs (32 cores / 64 threads) and 125\,GiB RAM.

For encrypted inference we replace every ReLU in the trained network with AppReLU (Section~\ref{subsec:apprelu}) and finetune the resulting model using the knowledge-distillation objective of Eq.~\ref{eq:kd_loss} for 200 epochs using the Adam optimizer (learning rate $10^{-4}$, weight decay $0$) with batch size $8{,}192$. The message-passing weights are frozen for a 10-epoch warm-up phase at $0.1\times$ the base learning rate; afterwards all parameters are trained jointly. Table~\ref{tab:kd_ablation} compares finetuning objectives for the $d=5$ AppReLU model averaged over 3 seeds on Small-HI. To account for different starting F1 scores from different seeds, we also report percentage of the original F1 preserved after finetuning. Knowledge Distillation outperforms the Na\"ive CE-only method by 9.97 F1 points on average and also has a substantially higher preservation of the baseline ReLU model.

\begin{table}[h]
  \centering
  \caption{AppReLU finetuning ablation on IBM AML Small-HI.}
  \label{tab:kd_ablation}
  \small
  \begin{tabular}{lrrr}
    \toprule
    Finetuning objective & F1 (\%) & Preservation (\%) \\
    \midrule
    CE-only              & 50.93 $\pm$ 2.24   & 78.31 $\pm$ 4.71 \\
    \rowcolor{rowhl}
    KD ($\alpha=0.7$, $T=3$, ours) & \textbf{60.90 $\pm$ 3.22} & \textbf{93.49 $\pm$ 3.34} \\
    \bottomrule
  \end{tabular}
\end{table}

\paragraph{Accuracy parity.}
Table ~\ref{tab:fhe_matchrate} reports plaintext F1 for the AppReLU-substituted, KD-finetuned model at $d \in \{1, 3, 5\}$ across all four splits. Since finetuning runs are computationally expensive, each $d \in \{1, 3, 5\}$ row reports a single finetuning run, initialized from one representative ReLU checkpoint. The reported numbers are compared to the baseline of Table ~\ref{tab:plaintext}'s mean results. Accuracy improves monotonically with $d$ on every split, since higher-degree compositions more closely approximate the discontinuous sign function, hence $d=5$ achieves the best score. We also note that the additional noise from more bootstrapping does not degrade the encrypted model's representational fidelity. Over a test set of 2000 seed edges (500 batches of 4), FHE and AppReLU $d=5$ plaintext predictions agree  99.25\% of the time.

Strikingly, the $d=5$ KD-finetuned model does not merely recover the ReLU baseline, but exceeds it. On Medium-HI, and Medium-LI, F1 exceeds the Table~\ref{tab:plaintext} means by 2.50 and 6.97 points respectively. The improvement on the Small splits is marginal by comparison. We hypothesize this reflects the amount of data available to the finetuning objective itself: AppReLU finetuning is effectively a second, distillation-regularized training phase, and its benefit should scale with dataset size, which is substantially larger for the Medium splits. 

% If this pattern holds beyond the single seed reported here, it is a favorable property for deployment, since real transaction networks are far larger than even the Medium IBM AML split — any accuracy gain that grows with data works in our favor at production scale.

% and the 15 mismatches are either borderline (9 cases with $|l_1-l_0| < 0.5$) or arise from floating-point overflow in the plaintext polynomial evaluation at boundary inputs (6 cases). 

% At $d=1$, the dedicated model achieves 99.95\% FHE match but its plaintext F1 of 37.71\% is too low for practical use. These results confirm $d=5$ as the minimum degree for both predictive accuracy and FHE reliability.

% \begin{table}[t]
%  \centering
%  \caption{FHE vs.\ AppReLU plaintext prediction match rate on Small-HI.
%           Each $d$ uses an independently distillation-trained model evaluated against
%           its own AppReLU plaintext forward pass.
%           ``Hard mismatches'' have plaintext decision margin $|l_1 - l_0| \ge 2.0$.}
%  \label{tab:fhe_matchrate}
%  \small
%  \resizebox{\columnwidth}{!}{%
%  \begin{tabular}{lcccc}
%    \toprule
%    Activation & Plaintext F1 (\%) & Seeds & Match rate (\%) & Hard mismatches \\
%    \midrule
%    $d=1$ & 37.71 & 2000 & 99.95 & 0 / 1 \phantom{0}(0\%) \\
%    $d=3$ & 53.73 & 2000 & 92.90 & 82 / 142 (58\%) \\
%    $d=5$ & \textbf{64.70} & 2000 & \textbf{99.25} & 0 / 15 \phantom{0}(0\%) \\
%    \bottomrule
%  \end{tabular}
%  }
%\end{table}

\begin{table}[t]
  \centering
  \resizebox{\columnwidth}{!}{
  \begin{threeparttable}
  \caption{AppReLU plaintext model F1 scores (\%) on all datasets.
           Each $d$ uses an independent distillation-trained model.}
  \label{tab:fhe_matchrate}
  \begin{tabular}{lcccc}
    \toprule
    Activation & Small-HI & Small-LI & Medium-HI & Medium-LI \\
    \midrule
    ReLU baseline & 63.72 & 30.12 & 63.20 & 24.52 \\
    \midrule
    $d=1$ & 37.71 & 13.36 & 46.72 & 8.67 \\
    $d=3$ & 53.73 & 27.55 & 65.49 & 29.86 \\
    $d=5$ & \textbf{64.70} & \textbf{31.40} & \textbf{65.70} & \textbf{31.49}\\
    \bottomrule
  \end{tabular}
  %\begin{tablenotes}
  % \item[1] We fix a \emph{single} seed as the base ReLU model and do a \emph{single} run of finetuning on top of it for this table. 
  %\end{tablenotes}
  \end{threeparttable}
  }
\end{table}

\paragraph{Inference latency.}
Table~\ref{tab:fhe_latency} reports end-to-end latency out of 100 sampled test batches at batch sizes  $b \in \{1, 2, 4, 8, 16, 32\}$, across all three polynomial degrees. Each call traverses one of two code paths depending on the sampled subgraph size $M'$: the \emph{single-ciphertext path} packs all edge features into one ciphertext when $M' \le 511$, and the \emph{block path} partitions into a grid of blocks otherwise. The reported figures exclude client-side encryption and decryption, key generation, network communication (Protocol steps 1, 3, and 5 in Section ~\ref{sec:inference_protocol}), and the row wise-to-bicyclic encoding transformation described in Section ~\ref{subsubsec:fhe_matmul}; our simulation encrypts edge features directly in bicyclic layout, whereas a deployed protocol would require the server to perform this transformation on receipt of each bank's row-wise ciphertexts.

At $d=5$ and batch size 1, the computations complete in $5.31$ seconds per inference. Latency rises only marginally as the batch size doubles to 2, since both configurations are resolved entirely within the single-ciphertext path. The trend changes at batch size 4, with a latency of $7.51$ seconds: 25\% of the subgraphs sampled with the 2-hop sampler already exceed the single-ciphertext capacity and are routed to the costlier block path. Latency spikes sharply beyond batch size 8, once the block path dominates: at $b=8$, 98\% of subgraphs require block partitioning, and from $b=16$ onward every sampled subgraph does. Table~\ref{tab:fhe_latency} also shows the cost of the rejected lower-degree alternatives: $d=5$ is always slower than $d=1$ ($2.5\times$ slower in the $b=1$ case), but we deem this cost justified by the substantial F1 gain.

To our knowledge, these figures are the first demonstration that encrypted GNN-based transaction classification is practical at this scale using a single consumer-grade GPU rather than a specialized accelerator or GPU cluster. We report that the overhead associated with our framework is consistent with the general computational cost associated with homomorphic evaluation, and thus view it not as a limitation to be eliminated but a price to be paid for obtaining strong accuracy and full feature confidentiality simultaneously. We also note that AML transaction screening is a compliance workflow, not a real-time authorization step, and a latency of a few seconds per query is comfortably within what such a workflow can absorb.
\begin{table}[t]
  \centering
  \caption{FHE inference latency (s) by batch size and App\-ReLU degree $d$,
           mean over 100 test batches on Small-HI. The row with best amortized latency ($\frac{time}{b}$ at $d=5$) is \colorbox{rowhl}{highlighted}.}
  %\resizebox{\columnwidth}{!}{
  \label{tab:fhe_latency}
  \begin{tabular}{crrrc}
    \toprule
    Batch size & $d=1$ & $d=3$ & $d=5$ & $d=5$ (amortized) \\
    \midrule
    1  & 2.14  & 3.74  & 5.31  & 5.31 \\
    2  & 2.37  & 3.96  & 5.56  & 2.78 \\
    4  & 3.50  & 5.54  & 7.51  & 1.88 \\
    8  & 6.84  & 10.20 & 13.56 & 1.69 \\
    \rowcolor{rowhl}
    16 & 15.16 & 20.98 & 26.42 & \textbf{1.65} \\
    32 & 43.66 & 54.56 & 64.15 & 2.00 \\
    \bottomrule
  \end{tabular}
  %}
\end{table}

\paragraph{Batch size, latency, and throughput trade-offs.}
The end-to-end latency of an inference call is the time to jointly classify all $b$ queries. To obtain the latency for one transaction, we divide the per-batch figures of Table~\ref{tab:fhe_latency} by $b$ and report it as amortized time in the last column. Dividing by batch size gives the best average per-transaction latency of 1.65 seconds with batch size 16. This distinction between end-to-end latency and per-transaction amortized time is operationally important. A bank that needs the fastest possible answer for one flagged transaction should request batch size 1: this minimizes the sampled subgraph, always triggering the faster single-ciphertext path, and finishes the fastest among all configurations. Conversely, a bank running scheduled bulk screening benefits from larger batches: the fixed costs incurred by homomorphic operations such as bootstrapping or matrix multiplications (Section~\ref{sec:fhe_methodology}) are amortized across every transaction rather than paid once per transaction, improving throughput.

The F1 impact of batch size is negligible: sweeping the plaintext batch size from 8,192 down to 4 moves Small-HI F1 only from $63.72 \pm 1.86$ to $63.74 \pm 1.61$, and Small-LI from $30.12 \pm 2.08$ to $31.16 \pm 1.36$, with every intermediate value within one standard deviation of both endpoints. Therefore, batch size can be tuned purely for latency and throughput without sacrificing accuracy. Batch size is not the only lever for controlling subgraph size: the neighborhood-sampling fan-out (Section ~\ref{sec:problem}) determines $M'$ directly and could be tuned instead of, or alongside, batch size. In summary, batch size 1 minimizes single-query latency for real-time or near-real-time scoring, while larger batches improve throughput for bulk or periodic screening, and a deployment can move between the two without retraining or affecting F1.

%% ================================================================
\section{Conclusion}
\label{sec:conclusion}

We addressed the problem of detecting money laundering across institutional boundaries, where the evidence is distributed over multiple banks that cannot share their transaction features under privacy regulations. We presented, to the best of our knowledge, the first framework for GNN inference carried out entirely on encrypted features in the AML setting: banks submit encrypted transaction features, while a server holding the global graph structure and model weights evaluates the network under CKKS. Two ideas make this possible: an incidence-matrix formulation of message passing and a composite minimax sign approximation in place of ReLU. On the IBM AML benchmark, the resulting model is competitive with plaintext baselines whose aggregation operations preclude efficient homomorphic evaluation, and its encrypted predictions achieve a competent accuracy at the cost of a homomorphic evaluation latency of several seconds per query, excluding communication and encryption overhead. We achieve an end-to-end latency of \textbf{5.31} seconds for a single query, amortizing to \textbf{1.65} seconds per transaction when batched for throughput. These results suggest that privacy regulation and collaborative cross-bank screening can coexist: encrypted GNN inference over transaction networks is practically achievable today.

\paragraph{Discussion.}  
Several directions remain open for extending this framework. System throughput can be increased straightforwardly with additional hardware, since the number of batches processed in parallel scales linearly with the number of GPUs; a consortium needing higher query volume can simply provision more GPUs. The latency of a single batch, however, is governed by a sequential chain of bootstrap operations that saturate one GPU and depend on each other. Hence, it cannot be reduced by parallelizing across GPUs alone, but a faster GPU can reduce it proportionally. Our formulation also supports encrypted account-level features as an additional input alongside transaction features (Algorithm~\ref{alg:incidencegnn-full}, line 10); since the IBM AML benchmark provides only transaction attributes, evaluating this extension, including its effect on detection performance and its homomorphic cost at realistic feature dimensions $d_v$, is a natural next step with suitable account-feature data.

\begin{acks}
  The support of the Swiss National Science Foundation (project number: 212158) for this work is gratefully acknowledged. This work was supported by the European Union’s Horizon Europe research and innovation program through the PRIVIDEMA Project under Grant 101167964. This work is funded by the European Union. Views and opinions expressed are however those of the author(s) only and do not necessarily reflect those of the European Union. Neither the European Union nor the granting authority can be held responsible for them.
\end{acks}

\bibliographystyle{ACM-Reference-Format}
\bibliography{sample-base}

@inproceedings{altman2023realistic,
  title={Realistic Synthetic Financial Transactions for Anti-Money Laundering Models},
  author={Erik Altman and Jovan Blanu{\v{s}}a and Luc Von Niederh{\"a}usern and Beni Egressy and Andreea Anghel and Kubilay Atasu},
  booktitle={NeurIPS Datasets and Benchmarks Track},
  year={2023}
}

@article{bicyclic2025,
  author={Chen, Jingwei and Yang, Linhan and Wu, Wenyuan and Liu, Yang and Feng, Yong},
  journal={IEEE Transactions on Information Forensics and Security},
  title={Homomorphic Matrix Operations Under Bicyclic Encoding},
  year={2025},
  volume={20},
  pages={1390--1404}
}

@inproceedings{cryptgnn2025,
author = {Sen, Pritam and Ma, Yao and Borcea, Cristian},
title = {CryptGNN: Enabling Secure Inference for Graph Neural Networks},
year = {2025},
isbn = {9798400715259},
publisher = {Association for Computing Machinery},
address = {New York, NY, USA},
url = {https://doi.org/10.1145/3719027.3765232},
doi = {10.1145/3719027.3765232},
booktitle = {Proceedings of the 2025 ACM SIGSAC Conference on Computer and Communications Security},
pages = {291–305},
numpages = {15},
location = {Taipei, Taiwan},
series = {CCS '25}
}

@phdthesis{gentry2009,
  author       = {Craig Gentry},
  title        = {A fully homomorphic encryption scheme},
  school       = {Stanford University},
  year         = {2009},
  note         = {\url{https://crypto.stanford.edu/craig/craig-thesis.pdf}}
}

@article{regev2013,
author = {Lyubashevsky, Vadim and Peikert, Chris and Regev, Oded},
title = {On Ideal Lattices and Learning with Errors over Rings},
year = {2013},
issue_date = {November 2013},
publisher = {Association for Computing Machinery},
address = {New York, NY, USA},
volume = {60},
number = {6},
issn = {0004-5411},
url = {https://doi.org/10.1145/2535925},
doi = {10.1145/2535925},
journal = {J. ACM},
month = nov,
articleno = {43},
numpages = {35}
}

@inproceedings {oblivgnn2024,
author = {Zhibo Xu and Shangqi Lai and Xiaoning Liu and Alsharif Abuadbba and Xingliang Yuan and Xun Yi},
title = {{OblivGNN}: Oblivious Inference on Transductive and Inductive Graph Neural Network},
booktitle = {33rd USENIX Security Symposium (USENIX Security 24)},
year = {2024},
isbn = {978-1-939133-44-1},
address = {Philadelphia, PA},
pages = {2209--2226},
url = {https://www.usenix.org/conference/usenixsecurity24/presentation/xu-zhibo},
publisher = {USENIX Association},
month = aug
}

@misc{tricycle2025,
  author={Lawrence Lim and Vikas Kalagi and Julia Novick and Jiaming Liu and Divyakant Agrawal and Amr El Abbadi},
  title={Tricycle: Private Transformer Inference with Tricyclic Encodings},
  howpublished={Cryptology ePrint Archive, Paper 2025/1200},
  year={2025}
}

@inproceedings{cheonetal2020,
  author={Cheon, Jung Hee and Kim, Dongwoo and Kim, Duhyeong},
  title={Efficient Homomorphic Comparison Methods with Optimal Complexity},
  booktitle={Advances in Cryptology -- ASIACRYPT},
  year={2020}
}

@misc{he2021fedgraphnn,
  title={FedGraphNN: A Federated Learning System and Benchmark for Graph Neural Networks},
  author={Chaoyang He and Keshav Balasubramanian and Emir Ceyani and Carl Yang and Han Xie and Lichao Sun and Lifang He and Liangwei Yang and Philip S. Yu and Yu Rong and Peilin Zhao and Junzhou Huang and Murali Annavaram and Salman Avestimehr},
  year={2021},
  eprint={2104.07145},
  archivePrefix={arXiv},
  primaryClass={cs.LG}
}

@inproceedings{slimbootstrapping,
  author={Bossuat, Jean-Philippe and Mouchet, Christian and Troncoso-Pastoriza, Juan and Hubaux, Jean-Pierre},
  title={Efficient Bootstrapping for Approximate Homomorphic Encryption with Non-sparse Keys},
  booktitle={Advances in Cryptology -- EUROCRYPT},
  year={2021}
}

@article{lee2022ppmlnn,
  author={Lee, Joon-Woo and Kang, Hyungchul and Lee, Yongwoo and Choi, Woosuk and Eom, Jieun and Deryabin, Maxim and Lee, Eunsang and Lee, Junghyun and Yoo, Donghoon and Kim, Young-Sik and No, Jong-Seon},
  journal={IEEE Access},
  title={Privacy-Preserving Machine Learning With Fully Homomorphic Encryption for Deep Neural Network},
  year={2022},
  volume={10},
  pages={30039--30054}
}

@article{Lee2021PreciseAO,
  title={Precise Approximation of Convolutional Neural Networks for Homomorphically Encrypted Data},
  author={Junghyun Lee and Eunsang Lee and Joon-Woo Lee and Yongjune Kim and Young-Sik Kim and Jong-Seon No},
  journal={IEEE Access},
  year={2021},
  volume={11},
  pages={62062--62076}
}

@misc{heongpu,
      author = {Ali Şah Özcan and Erkay Savaş},
      title = {{HEonGPU}: a {GPU}-based Fully Homomorphic Encryption Library 1.0},
      howpublished = {Cryptology {ePrint} Archive, Paper 2024/1543},
      year = {2024},
      url = {https://eprint.iacr.org/2024/1543}
}

@inproceedings{li2023lingcn,
  title={Lin{GCN}: Structural Linearized Graph Convolutional Network for Homomorphically Encrypted Inference},
  author={Hongwu Peng and Ran Ran and Yukui Luo and Jiahui Zhao and Shaoyi Huang and Kiran Thorat and Tong Geng and Chenghong Wang and Xiaolin Xu and Wujie Wen and Caiwen Ding},
  booktitle={Neural Information Processing Systems},
  year={2023}
}

@inproceedings{ran2022cryptogcn,
  author={Ran, Ran and Xu, Nuo and Wang, Wei and Quan, Gang and Yin, Jieming and Wen, Wujie},
  title={CryptoGCN: Fast and Scalable Homomorphically Encrypted Graph Convolutional Network Inference},
  booktitle={Neural Information Processing Systems},
  year={2022}
}

@inproceedings{ran2023penguin,
  title={Penguin: Parallel-Packed Homomorphic Encryption for Fast Graph Convolutional Network Inference},
  author={Ran Ran and Nuo Xu and Tao Liu and Wei Wang and Gang Quan and Wujie Wen},
  booktitle={Neural Information Processing Systems},
  year={2023}
}

@techreport{unodc2022,
  title       = {Estimating Illicit Financial Flows Resulting from Drug Trafficking and Other Transnational Organized Crimes},
  institution = {United Nations},
  author      = {{United Nations Office on Drugs and Crime (UNODC)}},
  year        = {2011},
}

@misc{weber2019anti,
  title={Anti-Money Laundering in Bitcoin: Experimenting with Graph Convolutional Networks for Financial Forensics},
  author={Mark Weber and Giacomo Domeniconi and Jie Chen and Daniel Karl I. Weidele and Claudio Bellei and Tom Robinson and Charles E. Leiserson},
  year={2019},
  eprint={1908.02591},
  archivePrefix={arXiv},
  primaryClass={cs.SI}
}

@inproceedings{yao2023fedgcn,
  author={Yao, Yuhang and Jin, Weizhao and Ravi, Srivatsan and Joe-Wong, Carlee},
  title={FedGCN: Convergence-Communication Tradeoffs in Federated Training of Graph Convolutional Networks},
  booktitle={Neural Information Processing Systems},
  year={2023}
}

@inproceedings{zhang2025ficgcn,
  title={Fic{GCN}: Unveiling the Homomorphic Encryption Efficiency from Irregular Graph Convolutional Networks},
  author={Zhaoxuan Kan and Husheng Han and shangyi shi and Tenghui Hua and Hang Lu and Xiaowei Li and Jianan Mu and Xing Hu},
  booktitle={International Conference on Machine Learning},
  year={2025}
}

@inproceedings{egressy2024provably,
  title={Provably Powerful Graph Neural Networks for Directed Multigraphs},
  author={Béni Egressy and Luc von Niederhäusern and Jovan Blanusa and Erik Altman and Roger Wattenhofer and Kubilay Atasu},
  booktitle={Proceedings of the AAAI Conference on Artificial Intelligence},
  year={2024}
}

@inproceedings{LaundroGraph,
  author={Cardoso, M\'{a}rio and Saleiro, Pedro and Bizarro, Pedro},
  title={LaundroGraph: Self-Supervised Graph Representation Learning for Anti-Money Laundering},
  booktitle={ACM International Conference on AI in Finance (ICAIF)},
  year={2022}
}

@article{tam2019identifying,
  title={Identifying Illicit Accounts in Large Scale E-payment Networks--A Graph Representation Learning Approach},
  author={Tam, Da Sun Handason and Lau, Wing Cheong and Hu, Bin and Ying, Qiu Fang and Chiu, Dah Ming and Liu, Hong},
  journal={arXiv:1906.05546},
  year={2019}
}

@inproceedings{gnn_fraud_rel,
  author={Dou, Yingtong and Liu, Zhiwei and Sun, Li and Deng, Yutong and Peng, Hao and Yu, Philip S.},
  title={Enhancing Graph Neural Network-based Fraud Detectors against Camouflaged Fraudsters},
  booktitle={CIKM},
  year={2020}
}

@misc{zhao2025design,
      title={DESIGN: Encrypted GNN Inference via Server-Side Input Graph Pruning},
      author={Kaixiang Zhao and Joseph Yousry Attalla and Qian Lou and Yushun Dong},
      year={2025},
      eprint={2507.05649},
      archivePrefix={arXiv},
      primaryClass={cs.CR},
}

@misc{megagnn,
  title={Multigraph Message Passing with Bi-Directional Multi-Edge Aggregations},
  author={H. Çağrı Bilgi and Lydia Y. Chen and Kubilay Atasu},
  year={2024},
  eprint={2412.00241},
  archivePrefix={arXiv},
  primaryClass={cs.LG}
}

@inproceedings{CKKS2017,
  author={Cheon, Jung Hee and Kim, Andrey and Kim, Miran and Song, Yongsoo},
  title={Homomorphic Encryption for Arithmetic of Approximate Numbers},
  booktitle={Advances in Cryptology -- ASIACRYPT},
  year={2017}
}

@article{FV2012,
  title={Somewhat Practical Fully Homomorphic Encryption},
  author={Junfeng Fan and F. Vercauteren},
  journal={IACR Cryptol. ePrint Arch.},
  year={2012},
  volume={2012},
  pages={144}
}

@misc{TFHE,
  Title   = {{TFHE}: Fast Fully Homomorphic Encryption Library},
  Author  = {Ilaria Chillotti and  Nicolas Gama and Mariya Georgieva and Malika Izabach{\`e}ne},
  Note    = {https://tfhe.github.io/tfhe/},
  Year    = {August 2016}
}

@inproceedings{mpnn,
  author={Gilmer, Justin and Schoenholz, Samuel S. and Riley, Patrick F. and Vinyals, Oriol and Dahl, George E.},
  title={Neural Message Passing for Quantum Chemistry},
  booktitle={International Conference on Machine Learning},
  year={2017}
}

@inproceedings{gcn,
  title={Semi-Supervised Classification with Graph Convolutional Networks},
  author={Thomas N. Kipf and Max Welling},
  booktitle={International Conference on Learning Representations},
  year={2017}
}

@inproceedings{graphsage,
  author={Hamilton, Will and Ying, Zhitao and Leskovec, Jure},
  title={Inductive Representation Learning on Large Graphs},
  booktitle={Neural Information Processing Systems},
  year={2017}
}

@inproceedings{Fey2019,
  title={Fast Graph Representation Learning with {PyTorch Geometric}},
  author={Fey, Matthias and Lenssen, Jan E.},
  booktitle={ICLR Workshop on Representation Learning on Graphs and Manifolds},
  year={2019},
}

@inproceedings{st_gcn,
  author={Yan, Sijie and Xiong, Yuanjun and Lin, Dahua},
  title={Spatial Temporal Graph Convolutional Networks for Skeleton-Based Action Recognition},
  booktitle={Proceedings of the AAAI Conference on Artificial Intelligence},
  year={2018}
}

@inproceedings{effendi2024aml,
  title={Privacy-Preserving Graph-Based Machine Learning with Fully Homomorphic Encryption for Collaborative Anti-Money Laundering},
  author={Effendi, Fabrianne and Chattopadhyay, Anupam},
  booktitle={International Conference on Security, Privacy, and Applied Cryptographic Engineering (SPACE)},
  year={2024}
}

@inproceedings{bnfolding,
  author={Jacob, Benoit and Kligys, Skirmantas and Chen, Bo and Zhu, Menglong and Tang, Matthew and Howard, Andrew and Adam, Hartwig and Kalenichenko, Dmitry},
  booktitle={IEEE/CVF Conference on Computer Vision and Pattern Recognition},
  title={Quantization and Training of Neural Networks for Efficient Integer-Arithmetic-Only Inference},
  year={2018}
}

@InProceedings{edge_dir,
  title = 	 {Edge Directionality Improves Learning on Heterophilic Graphs},
  author =       {Rossi, Emanuele and Charpentier, Bertrand and Giovanni, Francesco Di and Frasca, Fabrizio and G{\"u}nnemann, Stephan and Bronstein, Michael M.},
  booktitle = 	 {Proceedings of the Second Learning on Graphs Conference},
  pages = 	 {25:1--25:27},
  year = 	 {2024},
  editor = 	 {Villar, Soledad and Chamberlain, Benjamin},
  volume = 	 {231},
  series = 	 {Proceedings of Machine Learning Research},
}

% \appendix
% \input{chapters/A-MatMulOps}

\end{document}